\documentclass[%
reprint,
superscriptaddress,
amsmath,amssymb,
pra,
]{revtex4-1}

\usepackage{graphicx}
\usepackage{epstopdf}
\usepackage{epsfig}
\usepackage{float}
\usepackage{amsmath}
\usepackage[sort&compress]{natbib}

\newcommand{\ket}[1]{|#1\rangle}

\begin{document}

\preprint{APS/123-QED}

\title{Generating frequency-bin qubits via spontaneous four-wave mixing\\ in a photonic molecule}

\author{Ilya N. Chuprina}
\affiliation{Zavoisky Physical-Technical Institute, FRC Kazan Scientific Center of RAS, 10/7 Sibirsky Tract str., Kazan 420029, Russia}
\affiliation{Kazan Federal University, 18 Kremlyovskaya Str., Kazan 420008, Russia}
\author{Alexey A. Kalachev}
\email{a.a.kalachev@mail.ru}
\affiliation{Zavoisky Physical-Technical Institute, FRC Kazan Scientific Center of RAS, 10/7 Sibirsky Tract str., Kazan 420029, Russia}
\affiliation{Kazan Federal University, 18 Kremlyovskaya Str., Kazan 420008, Russia}

\date{\today}

\begin{abstract}
A scheme for heralded generation of frequency-bin photonic qubits via spontaneous four-wave mixing in a system of coupled microring resonators (photonic molecule) is developed so that the qubit state is fully controlled by the frequency amplitude of the pump field. It is shown that coupling parameters of the photonic molecule can be optimized to provide generation of nearly pure heralded photons. The heralded qubit state does not depend on the microring resonator mode frequency, thereby allowing to take advantage of frequency multiplexing for improving efficiency of quantum information processing.
\end{abstract}

\pacs{42.50.Dv,42.60.Da}

\keywords{spontaneous four-wave mixing; microring resonator; frequency-bin qubit; heralded single-photon source}
\maketitle


\section{\label{sec:intro}Introduction}
Heralded single-photon sources continue to gain attention as a basic ingredient of optical quantum technologies \cite{Eisaman:2011cc,Caspani:2017gq}. In particular, preparing single-photon states via spontaneous four-wave mixing (SFWM) in microring resonators is a promising approach to develop compact and effective on-chip devices that is compatible with existing CMOS technology. In the SFWM process, two pump laser photons inside a third-order nonlinear optical material can spontaneously convert into a pair of photons, so that one of them can be used to herald the presence of the other. In doing so, microring resonators have proved to be promising structures for enhancing optical fields and creating effective quantum light sources \cite{Caspani:2017gq}. In this respect, significant experimental progress has been achieved in demonstration of high efficiency of the nonlinear process \cite{Gaeta:2008bc,Azzini:2012io,Savanier:2016kb}, and narrow bandwidth \cite{Reimer:2014ev} and high purity \cite{Faruque:2018ec} of the emitted photons. In addition, near deterministic emission can potentially be achieved using multiplexing techniques \cite{Collins:2013eu,Christensen:2015dn,Mendoza:2016jz,Xiong:2016bv,Mosley:2016gr,Puigibert:2017fh,Joshi:2018dj}, which is expected to be quite efficient when using photon number resolving detectors \cite{Okamoto:2016ee}. This approach employs multiple heralded single-photon sources so that whenever one of them signals the successful heralding, the corresponding photon is routed to the output. In such a way, high total efficiency can be achieved without increasing the pump power for each source, thereby mainaining quality of the emitted single-photon state.

The use of photons as quantum information carriers assumes developing effective methods of controlling their basic properties that are typically used for encoding such as temporal waveform, orbital angular momentum and polarization \cite{Flamini:2019bp}. Recently, there has been a growing interest in using frequency-bin photonic qubits for quantum information processing \cite{Lukens:2017go,Arie:2018el,Lu:2018br,Lu:2019fp}. In this case, a qubit state is encoded in a superposition of two single-photon states with different frequencies. Similar to time-bin qubits, where superposition of two single-photon wave packets separated in time is used for encoding, they demonstrate high stability against environmental fluctuations when transmitted via a single spatial channel, but do not require stabilisation of optical interferometric schemes for detection or fast electronics for processing. For the frequency-bin qubit to be prepared from a single-frequency input pulse, one can take advantage of electro-optic phase modulators and Fourier-transform pulse shapers \cite{Lu:2018br,Lu:2019fp} since the problem is equivalent to implementing an arbitrary single-qubit gate in the frequency domain. In particular, such a single-qubit gate may be placed at the output of a heralded single-photon source. However, the resulting heralding efficiency (the probability of the photonic qubit generation when the heralding signal appears) may be significantly reduced by insertion losses and less than unit success probability of the gate. 

In the present paper, we develop a scheme for generating frequency-bin single-photon qubits via SFWM in a system of coupled micro-resonators which is often referred to as a photonic molecule \cite{Bayer:1998db,Rakovich:2010gc,Li:2017ck,Zhang:2019ic}. The strong coupling between micro-resonators provides frequency splitting of the modes \cite{Li:10,bogaerts2012silicon,Miller:15} that can be used for preparing narrowband frequency-bin qubits compatible with quantum memory devices relying on atomic transitions. The developed scheme allows one to prepare photonic qubits by controlling the frequency amplitude of the pump field, thereby avoiding additional losses introduced by external modulators or pulse shapers and achieving the highest possible heralding efficiency. In addition, the proposed system of coupled microring resonators makes it possible to generate pure single-photon states (transform-limited single-photon wave packets) \cite{Chuprina:2018dk}. The resulting state does not depend on the resonator mode frequency, which allows one to take advantage of scalable frequency multiplexing schemes \cite{Puigibert:2017fh,Joshi:2018dj} for approaching deterministic emission of indistinguishable photons. Moreover, since the frequency-bin qubit states can be processed without reference to optical frequency, high generation rates of these states may be attainable with no need for optical frequency conversion.  

\section{\label{sec:model}Basic equations}

\subsection{\label{sec:molecule}The model}

Let us consider a system of four microring resonators such that one of them (central) is coupled to strait waveguides (buses) through the others [Fig.~1(a)]. It is assumed that the SFWM process occurs in the central ring, while other rings are used for loading the pump field and unloading the generated photons. This can be achieved by choosing appropriate sizes of the resonators and tuning their frequencies. In particular, when the free spectral range of the outer rings is three times smaller than that of the central ring, it is possible to tune them so that the signal and idler photons either go to different waveguides without the pump field or go to the pump waveguide, which depends on their frequency [Fig.~1(b)]. In what follows, we are interested in the signal and idler fields that are demultiplexed into the different strait waveguides.

\begin{figure}[ht]
    \includegraphics[width=0.45\textwidth]{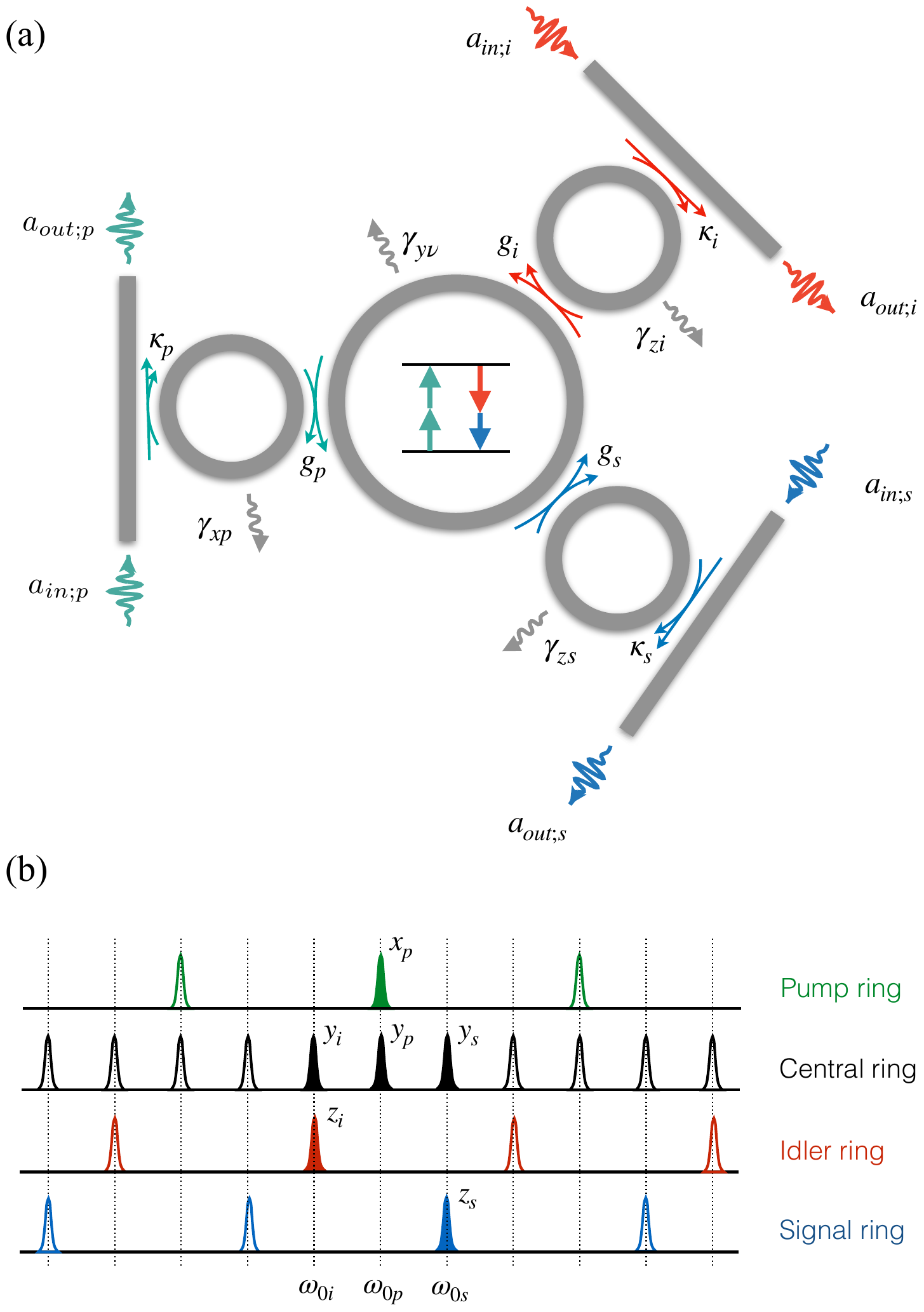}
    \caption{General scheme of the photonic molecule (a) and an example of the resonator tuning for frequency demultilexing the emitted photons (b)}
	\label{fig:scheme}
\end{figure}

The fields in the microring resonators are described in the frequency domain by the annihilation operators $x_\nu(\omega)$, $y_\nu(\omega)$ and $z_\nu(\omega)$ ($\nu=\{i,s,p\}$), so that $x_p(\omega)$ corresponds to the pump field in the left outer ring, $y_p(\omega)$, $y_s(\omega)$, and $y_i(\omega)$ --- to the pump, signal  and idler fields, respectively, in the central ring, and $z_s(\omega)$, $z_i(\omega)$ --- to the signal and idler fields, respectively, in the right outer rings. The central frequencies of the modes in different microrings are assumed to be matched with each other as shown in Fig.~1(b). The input and output fields in the strait waveguides are described by the annihilation operators $a_{in;\nu}(\omega)$ and $a_{out;\nu}(\omega)$. The coupling parameters between the microrings are denoted by $g_\nu$, while those between the microrings and waveguides are labelled $\kappa_\nu$. Finally, linear losses in the microring resonators are described by the rates $\gamma_{q\nu}$ ($q=\{x,y,z\}$). 

\subsection{\label{sec:molecule}Input-output relations}

A key point of the model description is the input-output relations between the resonator and waveguide fields. For the lossy resonators they can be derived from the Heisenberg--Langevin equations by introducing phantom waveguides (see Appendix), which gives the following result:
\begin{eqnarray}\label{in-out_n}
& y_\nu(\omega)&=\frac{4g_\nu\sqrt{\kappa_\nu}\,a_{in;\nu}(\omega)}{-4\Delta_{\nu}^2+4g_\nu^2+2i\Delta_{\nu}(\kappa_\nu+2\gamma_{\nu})+\kappa_\nu \gamma_\nu+\gamma_\nu^2},\nonumber\\
& &\equiv M_\nu(\omega) a_{in;\nu}(\omega),
\end{eqnarray}
where $\Delta_\nu=\omega_{0\nu}-\omega$, and $\omega_{0\nu}$ is the central frequency of the resonator mode corresponding to the $\nu$th field. For simplicity we assume here that $\forall q$ $\gamma_{q\nu}=\gamma_\nu$ (general expressions are presented in Appendix). To express the cavity field operators $y_\nu$ in terms of the output fields $a_{out;\nu}(\omega)$, $M_\nu(\omega)$ in Eqs.~(\ref{in-out_n}) should be replaced by $M_\nu^\ast(\omega)$. In what follows, the functions $M_\nu(\omega)$ are referred to as transfer functions describing propagation of optical signals at frequency $\omega$ from the central microring resonator to the strait waveguides.

\subsection{\label{sec:biphoton}Biphoton state}

To calculate the state of the biphoton field, we can take advantage of the first-order perturbation theory of the cavity-assisted SFWM (see, e.g., \cite{Chen:11,GarayPalmett:2012bv,Chuprina2017}). The effective SFWM Hamiltonian can be written as
\begin{equation} \label{H_SFWM}
\mathcal{H}_{SFWM}(t) = \zeta y_p(t)y_p(t)y_s^\dagger(t)y_i^\dagger(t),
\end{equation}
where $\zeta$ is the effective nonlinearity that takes into account $\chi^{(3)}$ of the nonlinear material, mode overlapping and other parameters. By applying the first--order perturbation theory, the state vector of the generated biphoton field is calculated as 
\begin{eqnarray}
& \nonumber \ket{\psi} = &\ket{0}\ket{\alpha} - \frac{i\zeta}{\hbar (2\pi)^2} \int dt d\omega_p d\omega_i d\omega_s\\
& &\times y_p(\omega_p)y_p(\omega_p)y^\dagger_{i}(\omega_i)y^\dagger_{s}(\omega_s)\,e^{i\Delta\omega t}\ket{0}\ket{\alpha},
\end{eqnarray}
where $\ket{0}=\ket{0_s}\ket{0_i}$ is the vacuum state of the signal and idler fields, $\ket{\alpha}$ is the coherent state of the pump field with a complex amplitude $\alpha$ (so that $y_p(t)\ket{\alpha}=\alpha(t)\ket{\alpha}$), and $\Delta\omega=2\omega_p-\omega_i-\omega_s$ is the frequency mismatch. From the input--output relations (\ref{in-out_n}) we obtain
\begin{eqnarray}\label{eq:state}
& \nonumber \ket{\psi} = &\ket{0}\ket{\alpha} - \frac{i\zeta}{\hbar \sqrt{2\pi}^3} \int d\omega_i d\omega_s \mathcal{F}(\omega_i,\omega_s)\\
& &\times y^\dagger_{out;i}(\omega_i)y^\dagger_{out;s}(\omega_s)\ket{0}
 \ket{\alpha},
\end{eqnarray}
where 
\begin{equation}\label{JSA}
\mathcal{F}(\omega_i,\omega_s) = \mathcal{I}_p(\omega_i,\omega_s) M_i(\omega_i) M_s(\omega_s)
\end{equation}
is the joint spectral amplitude (JSA) of the biphoton field, and
\begin{eqnarray}\label{eq:convolv}
&   \mathcal{I}_p(\omega_i,\omega_s) = &\int d\omega_p M_p(\omega_s+\omega_i-\omega_p)M_p(\omega_p)\nonumber \\
&    &\times \alpha(\omega_s+\omega_i-\omega_p)\alpha(\omega_p)
\end{eqnarray}
is the convolution of the spectral amplitude of the pump field in the central resonator. 

To analyze the spectral correlations between the signal and idler photons one can use the Schmidt decomposition of the JSA:
\begin{equation}\label{eq:Schmidt decomposition}
    \mathcal{F}(\omega_i,\omega_s)=\sum_n\sqrt{\lambda_n}\,\psi_n(\omega_i)\phi_n(\omega_s),
\end{equation}
where the Schmidt coefficients satisfy the condition $\sum_n\lambda_n=1$. The Schmidt decomposition provides a convenient measure of entanglement within a pure state, $K = 1/\sum_n\lambda_n^2$, which is called Schmidt number. If the state is entangled, then there is more than one term present in Eq.~(\ref{eq:Schmidt decomposition}), and $K>1$. The purity of the heralded single-photon states is proportional to the inverse degree of entanglement of the total state and calculated as $P=1/K$.  In the case of a separable biphoton state, there is only one non-vanishing Schmidt coefficient $\lambda = 1$ yielding minimum Schmidt number $K=1$ and maximum purity $P=1$. To illustrate the spectral correlations between the emitted photons it is convenient to use the joint spectral intensity $|\mathcal{F}(\omega_i,\omega_s)|^2$ that can be understood as the two-dimensional probability distribution of the signal and idler photons in the frequency domain.

\section{\label{sec:qubit_state}Frequency-bin qubits}

\subsection{Strong coupling}

Let us consider the case of strong coupling between the central and outer resonators for the signal and idler fields: $g_\mu>\kappa_\mu$ ($\mu=\{i, s\}$). In this case, the transfer functions $M_\mu(\omega_\mu)$ take the form of two Lorentzian-like peaks:
\begin{equation}
M_\mu(\omega_\mu)=M_{\mu}^{-}(\omega_\mu)+M_{\mu}^{+}(\omega_\mu),	
\end{equation}
separated in frequency by 
\begin{equation}
\delta_\mu=2\sqrt{g_\mu^2-\kappa_\mu^2/8},
\end{equation}
provided that $\kappa_\mu<\sqrt{8}\,g_\mu$. Here $M_{\mu}^{\pm}(\omega_\mu)$ stands for the mode profile around the central frequency $\omega_{0\mu}^{\pm}=\omega_{0\mu}\pm\delta_\mu/2$, which is characterized by spectral width (FWHM) of $\kappa_\mu+\gamma_{y\mu}+\gamma_{z\mu}$ in the case of small linear losses. The components of the double line correspond to symmetric and antisymmetric states of the fields distributed between the central and outer resonators. Due to selective resonant coupling, the pump field penetrates only in the central ring, where SFWM occurs. However, the signal and idler photons are created in the collective modes of the corresponding branches of the photonic molecular. 

\begin{figure}[ht]
    \includegraphics[width=0.4\textwidth]{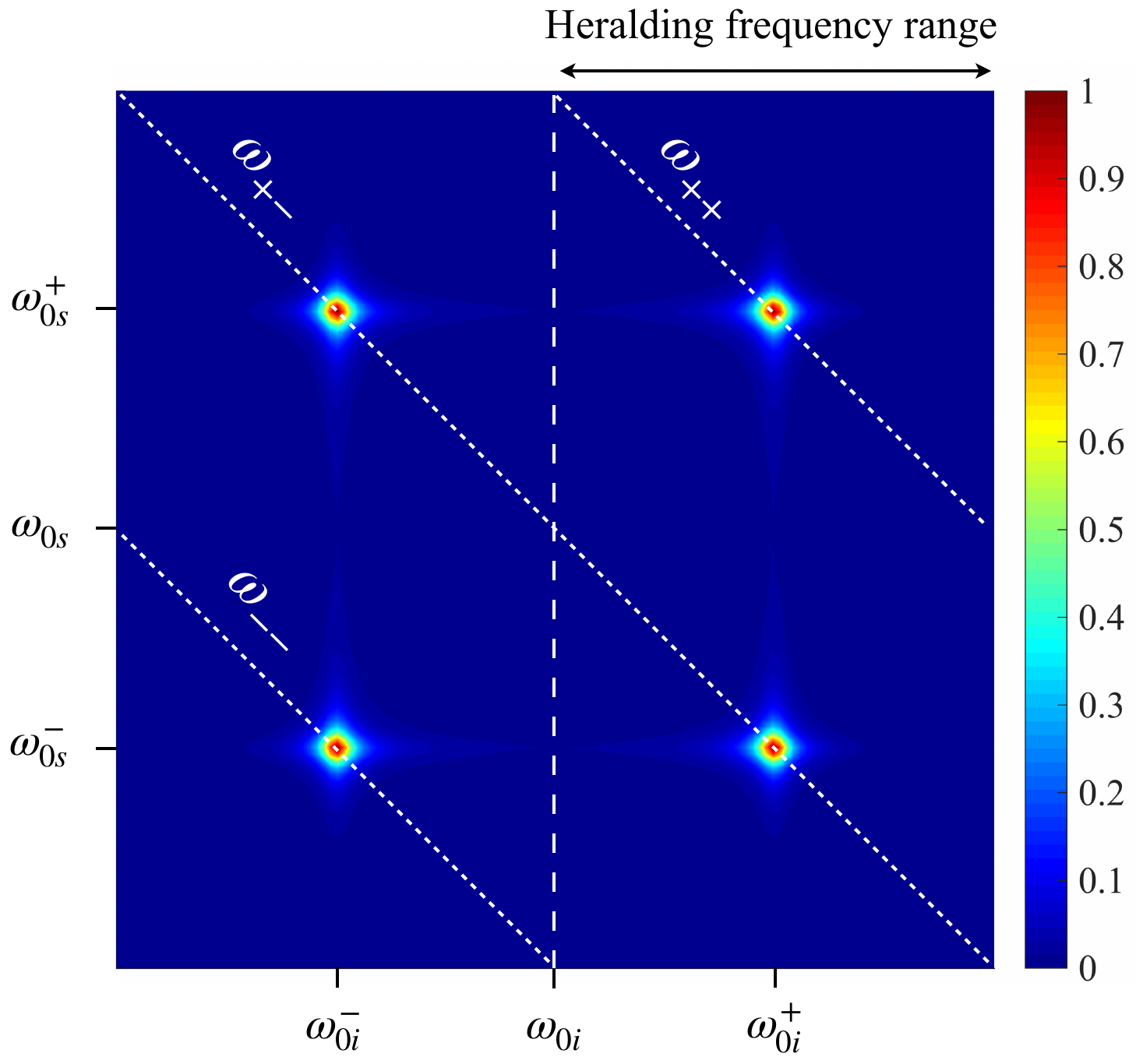}
    \caption{Joint spectral intensity, $|\mathcal{F}(\omega_i,\omega_s)|^2$, for the case of $g_\mu\gg\kappa_\mu$, $\gamma_\mu/g_\mu\ll 1$, and an infinitively broad spectral amplitude of the pump field.}
	\label{fig:double0}
\end{figure}
The frequency splitting of the signal and idler modes results in appearing four peaks in the JSA, if the pump field is broad enough, which correspond to possible combinations of the photon frequencies (Fig.~\ref{fig:double0}). These peaks form three groups: i) both photons have smaller frequencies ($\omega_{0i}^-+\omega_{0s}^-=2\omega_{--}$), ii) both photons have larger frequencies ($\omega_{0i}^++\omega_{0s}^+=2\omega_{++}$), and iii) one photon has smaller while another has larger frequency ($\omega_{0i}^++\omega_{0s}^-=\omega_{0i}^-+\omega_{0s}^+=2\omega_{+-}$). By making the pump field narrower, it is possible to excite only one of these groups. On the other hand, a broadband pump field allows one to control the relative amplitudes and phases of these groups via manipulating its spectral amplitude. It is the approach that is used here for frequency-bin preparation.   

\subsection{\label{sec:conv}Preparing a qubit state}

Let the pump field spectral amplitude consists of two Gaussian components with the central frequencies $\omega_{+-}$ and $\omega_{++}$:
 \begin{eqnarray}\label{eq:Pump state}
 &   \alpha(\omega_p) \sim &\sqrt{A}\,e^{-\sigma(\omega_p-\omega_{+-})^2-i\phi_a/2}\nonumber\\ 
 & &+\sqrt{B}\,e^{-\sigma(\omega_p-\omega_{++})^2-i\phi_b/2},
 \end{eqnarray}
each of the same spectral bandwidth $\Delta\omega_p=\sqrt{2\ln 2/\sigma}$ that is small compared to the mode splitting $\delta_\mu$. For such a pulse to be applied as a pump field without distortion, the resonator mode for the pump field should be sufficiently broad and symmetric with respect to $\omega_{+-}$ and $\omega_{++}$. Let us consider the case of $\kappa_p<\sqrt{8}\,g_p$, when such a mode consists of two components with the frequencies $\omega_{0p}^{\pm}=\omega_{0p}\pm\delta_p/2$. They match the signal and idler modes provided that $\omega_{0p}=(\omega_{+-}+\omega_{++})/2$ and $\delta_p=\delta_\mu$. The latter condition can be satisfied for arbitrary large values of $\kappa_p$ with respect to $\kappa_\mu$. In this case, a broadband pump field with respect to the signal and idler fields can be applied, thereby achieving high spectral purity of the emitted photons \cite{Vernon:17,Chuprina:2018dk}. 

If $\kappa_p, \Delta\omega_p\gg\kappa_{\mu}$, we can take $M_p(\omega_p)$ and $\alpha(\omega_p)$ out of the integrals in Eq.~(\ref{eq:state}) as slowly varying functions with respect to $M_\mu(\omega_\mu)$ for each peak of JSA, which gives 
 \begin{eqnarray}
 &   \nonumber \mathcal{F} (\omega_i, \omega_s) \sim &A\,e^{-i\phi_a} M_{i}^{-}(\omega_i) M_{s}^{+}(\omega_s)\\
 &   \nonumber &+A\,e^{-i\phi_a} M_{i}^{+}(\omega_i) M_{s}^{-}(\omega_s)\\
 &   &+B\,e^{-i\phi_b} M_{i}^{+}(\omega_i) M_{s}^{+}(\omega_s).
    \label{eq:copmF}
 \end{eqnarray}
The forth peak, which is proportional to $M_{i}^{-}(\omega_i) M_{s}^{-}(\omega_s)$, may be ignored if $\Delta\omega_p$ is small compared to $\delta_\mu$. Now, detection of the idler photon at the higher frequency (i.e., selection of the idler photon component proportional to $M_i^{+}(\omega_i)$ as shown in Fig.~\ref{fig:double0}) creates superposition of two frequency components for the signal field:
 \begin{equation}\label{eq:Qubit state}
    \ket{\psi_{s}} = A\,e^{-i\phi_a}\ket{a}+B\,e^{-i\phi_b}\ket{b},
 \end{equation}
 where
 \begin{eqnarray}
 &\ket{a} &= \int d\omega_s M_{s}^{-}(\omega_s)y^\dagger_{out; s}(\omega_s)\ket{0},\\
 &\ket{b} &= \int d\omega_s M_{s}^{+}(\omega_s)y^\dagger_{out; s}(\omega_s)\ket{0},
 \end{eqnarray}
 and $A$ and $B$ should be normalized so that $\langle\psi_s|\psi_s\rangle=1$. As a result, the angles $\theta=\arctan(A/B)$ and $\phi=\phi_b-\phi_a$ may be considered as those defining orientation of the Bloch vector representing the qubit state. The qubit state amplitudes in Eq.~(\ref{eq:Qubit state}) are simply squares of the pump field amplitudes in Eq. (\ref{eq:Pump state}) provided that crosstalk between the frequency bins of the pump field is negligible.
 
\begin{figure}[ht]
    \includegraphics[width=0.5\textwidth]{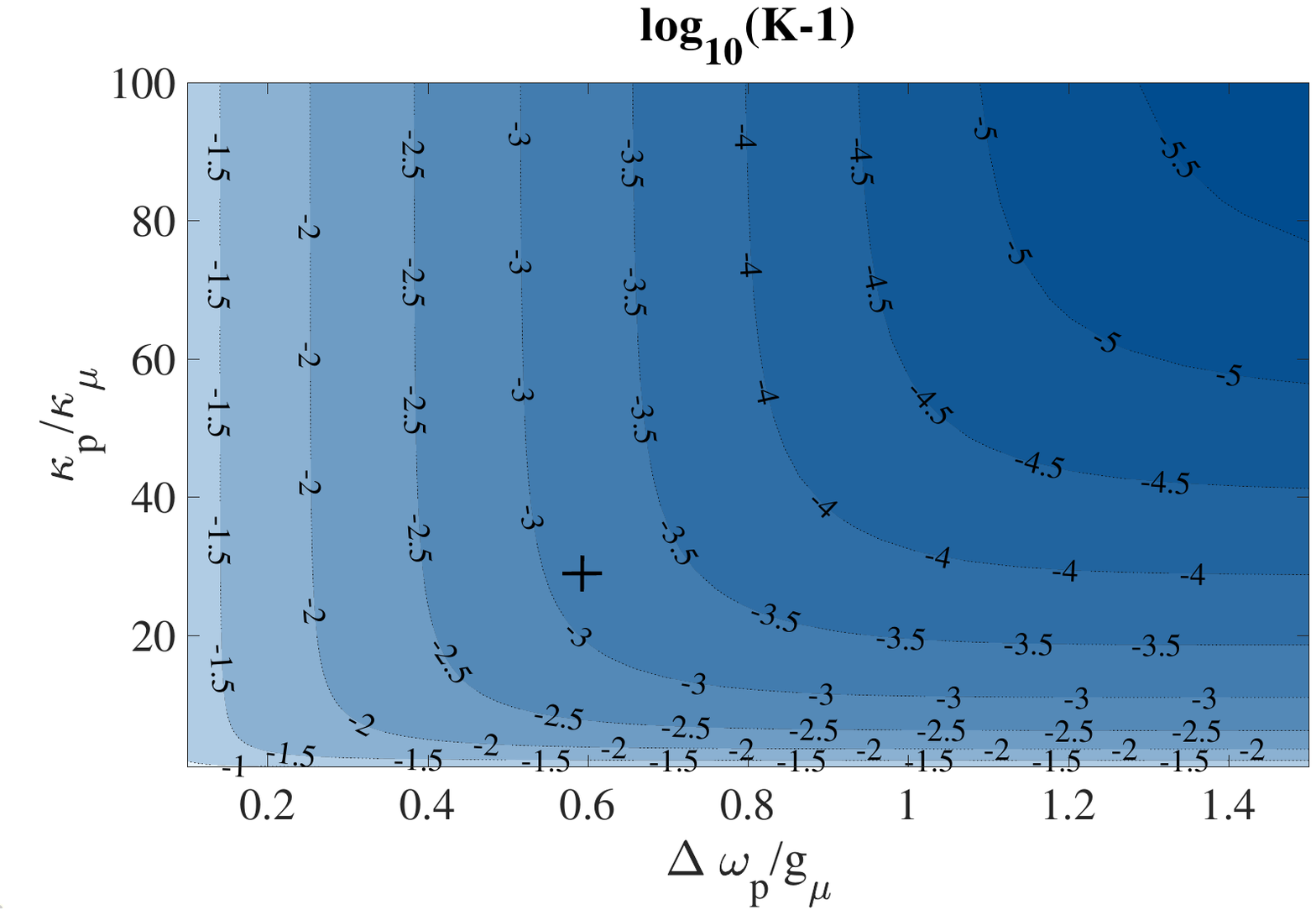}
    \caption{The Schmidt number $\log_{10}(K-1)$ as a function of the resonator linewidth for the pump field $\kappa_p$ and spectral width of the pump field $\Delta \omega_p$, normalized to the coupling constants $\kappa_\mu$ and $g_\mu$, respectively. The calculations are made for the case of $\delta_p=\delta_\mu$, $\kappa_\mu/g_\mu=0.1$ and $\gamma_\nu/g_\nu \ll 1$. The cross corresponds to the values $\kappa_p/\kappa_\mu=30$ and $\Delta\omega_p/g_\mu=0.6$.}
	\label{fig:Shcmidtmap}
\end{figure}

\begin{figure}[ht]
    \includegraphics[width=0.5\textwidth]{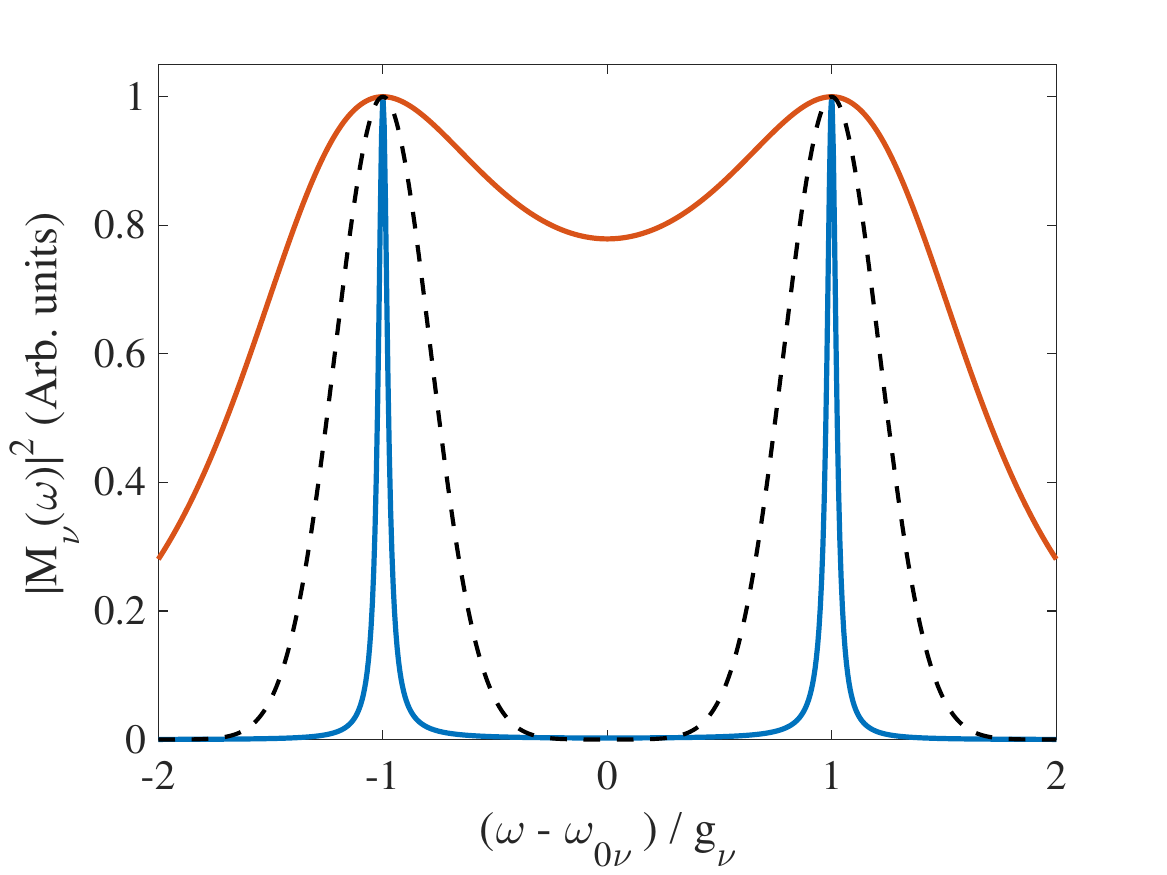}
    \caption{Transfer function profiles for the pump field (red solid curve) and signal/idler field (blue solid curve) for the following values of parameters: $\kappa_p/\kappa_\mu=30$, $\kappa_p/g_p=2.06$, $\kappa_\mu/g_\mu=0.1$ and $g_p/g_\mu=1.46$. Dashed curve illustrates spectral amplitude of the pump field for the case of $\Delta\omega_p/g_\mu=0.6$.}
	\label{fig:Spectra}
\end{figure}

Thus frequency-bin photonic qubit is conditionally prepared such that its state is totally controlled by the spectral amplitude of the pump field. This is possible provided that spectral width of the signal (heralded) mode is much smaller than that of the pump field components, which in turn should be smaller that the resonator linewidth. To illustrate these conditions we calculated the Schmidt number of the qubit frequency bins as a function of the spectral width of the pump field $\Delta\omega_p$ and resonator linewidth for the pump field, which for small losses is reduced to $\kappa_p$ (Fig.~\ref{fig:Shcmidtmap}). 
It can be seen that for each value of the Schmidt number the contour line is almost independent on the spectral width of the pump field if the latter exceeds some minimum value. The same is true for the dependence of the Schmidt number on the resonator linewidth for the pump field. Therefore, we can choose the optimal combination of $\kappa_p$ and $\Delta\omega_p$ so that both of them be close to their minimal values. Minimizing the ratio $\Delta\omega_p/g_p$ allows one to reduce crosstalk between two frequency bins of the pump field, while smaller values of $\kappa_p/\kappa_\nu$ require a smaller difference between the coupling constants, which may be useful for avoiding additional losses in the case of small gap widths \cite{Tseng:13}. As an example, Fig.~\ref{fig:Spectra} shows transfer functions and spectral amplitude of the pump field for the point $\kappa_p/\kappa_\mu=30$ and $\Delta\omega_p/g_\mu=0.6$ (marked by the cross in Fig.~3), corresponding to the Schmidt number $K$ of about 1.001. 

\section{Implementation issues}
As an example of a promising CMOS-compatible platform for implementing the proposed scheme, we consider microring resonators formed by silicon-nitride based strip waveguides ($\text{Si}_3\text{N}_4/\text{SiO}_2$), which have been successfully used in integrated nonlinear photonics \cite{Moss:2013kv}. In particular, recent achievements include the fabrication of ultra-low-loss waveguides providing the microring resonator quality factor exceeding $10^7$ \cite{Pfeiffer:2018cd}, dispersion engineering \cite{Herkommer:2017cp}, and generation of entangled photon states \cite{Weiner:2018jk,Lu:2019dd,Martin:2019hl}. For the multimode scheme shown in Fig.~1 to be realised, the pump field should correspond to a resonator mode in a zero-dispersion region of the central microring resonator so that the signal and idler photons can be emitted into the adjacent modes that are separated from the pump mode by equal frequency intervals (the group velocity dispersion is negligible). This condition can be fulfilled in the telecommunication C-band if the waveguides are sufficiently thick \cite{Kruckel:2015hv,Chuprina2017}. In doing so, a quantum frequency comb containing $\sim 40$ entangled modes with linewidths of $\sim 100$~MHz and frequency separations $\sim 50$~GHz has been recently generated \cite{Weiner:2018jk}. Such spectral characteristics, which seem to be the most appropriate for the present scheme, can be achieved by using microresonators of about 500 $\mu$m in radius with a loaded quality factor of $\sim 10^6$. Line-by-line pulse shaping of optical pump pulses with sub-GHz spectral resolution \cite{Willits:12} allows one to use $\sim 1$~GHz frequency splitting of the resonator modes for preparing narrow-band frequency-bin qubits. In addition, for the frequency combs of different resonators to be perfectly matched, as shown in Fig.~1(b), the free spectral range of them should be tuned with an accuracy better than the resonator linewidth. This requirement can be satisfied by combining two approaches: (i) fabrication imperfections can be compensated by post deposition trimming, which reduces the standard deviation of resonance wavelength to a value of several GHz \cite{Fan:18}; (ii) more accurate tuning and stabilization can be achieved by the use of integrated microheaters providing a resonance wavelength accuracy of several MHz \cite{Miller:15}. 

Finally, it should be noted that high heralding efficiency can be achieved only in the case of an over-coupled system of the microring resonators and strait waveguides \cite{Vernon:2016iv}, which means that the linear losses in the rings $\gamma_{q\mu}$ should be much smaller than the coupling rates with the output channels $\kappa_\mu$. To be more precise, following \cite{Vernon:2016iv} the heralding efficiency can be estimated in this case as $\eta=\kappa_\mu/(\kappa_\mu+\gamma_{y\mu}+\gamma_{z\mu})$, where the index $\mu=\{i,s\}$ corresponds to the heralded field. The coupling constants between the microrings $g_\mu$ do not contribute to the photon losses and therefore to the heralding efficiency. On the other hand, as shown in \cite{Vernon:2016iv}, there is a trade-off between the heralding efficiency and the photon pair generation rate under a certain level of the pump power. To generate photons at the same rate with a higher heralding efficiency, the pump power needs to be increased. In this context, addressing other factors that reduce the heralding efficiency, which was done in the present work, is important for developing efficient microresonator frequency-bin qubit sources.

\section{Conclusion}

We have demonstrated how to conditionally prepare photonic frequency-bin qubits via spontaneous four-wave mixing in a system of coupled microring resonators. The developed scheme provides highest possible heralding efficiency since the resulting qubit state is fully controlled by the frequency amplitude of the pump field. We have proposed to use double resonant lines of the coupled microring resonators for the frequency encoding, which allows one to prepare narrowband frequency-bin qubits independent on the mode frequency. This can be exploited to enhance qubit generation rate via frequency multiplexing to the extent possible. The proposed approach can potentially be generalized for more complicated photonic molecules, allowing generation of high-dimensional photonic states. It can also be adapted to heralded single-photon sources based on spontaneous parametric down-conversion in second-order nonlinear microresonators.

\section*{Acknowledgements}
The work was supported by the Russian Science Foundation (project No. 16-12-00045).

\appendix*
\renewcommand\thefigure{A.\arabic{figure}}
\section{}
To derive the input-output relations (\ref{in-out_n}) for the system of coupled microring resonators, we use the Heisenberg--Langevin approach. The Hamiltonian of the system can be written as
\begin{eqnarray}\label{eq:Hamilt}
&\nonumber \mathcal{H} = &\mathcal{H}_\text{sys}+\mathcal{H}_\text{bath}+\mathcal{H}^\text{sys}_\text{int}+\mathcal{H}^\text{bath}_\text{int}\\ 
& &+\mathcal{H}_\text{phantom}+\mathcal{H}^\text{phantom}_\text{int},
\end{eqnarray}
where
\begin{eqnarray}\label{exten_hamilt2}
&\nonumber	\mathcal{H}_\text{sys} = &\hbar\omega_{0x,p}\, x_p^\dagger x_p + \sum_{\nu=p,i,s} \hbar\omega_{0y,\nu}\,y_\nu^\dagger y_\nu \\
&	&+ \sum_{\mu={i,s}} \hbar\omega_{0z,\mu}\, z_\mu^\dagger z_\mu
\end{eqnarray}
is the free-field Hamiltonian for the resonators,
\begin{equation}
\mathcal{H}_\text{bath} = \int d\omega\,\hbar\omega \sum_{\nu=i,p,s} a_\nu^\dagger(\omega) a_\nu(\omega)
\end{equation}
is the external bath Hamiltonian,
\begin{equation}
\mathcal{H}_\text{int}^\text{sys} = i\hbar g_p x_p^\dagger y_p+i \hbar g_i z_i^\dagger y_i + i \hbar g_s z_s^\dagger y_s+\text{H.c.}
\end{equation}
is the coupling between the rings, and
\begin{eqnarray}
&\nonumber	\mathcal{H}_\text{int}^\text{bath} = &\frac{i\hbar}{\sqrt{2\pi}} \int d\omega\, \big[ \sqrt{\kappa_p}\, x_p^\dagger a_p(\omega) +
	\sqrt{\kappa_i}\,z_i^\dagger a_i(\omega)\\
&	&+\sqrt{\kappa_s}\,z_s^\dagger a_s(\omega)+\text{H.c.}\big]
\end{eqnarray}
is the coupling between the microring resonators and strait waveguides. To account for the linear losses in the microring resonators we take advantage of effective phantom waveguides, to which each ring linearly couples (Fig.~\ref{fig:scheme}).
\setcounter{figure}{0}  
\begin{figure}[ht]
    \includegraphics[width=0.4\textwidth]{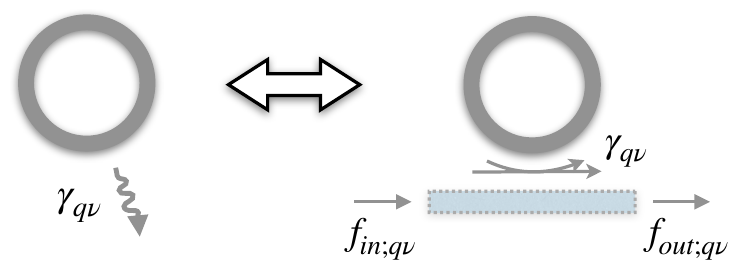}
   \caption{Illustrating a phantom waveguide (on the right) simulating the linear losses in the microring resonator (on the left)}
	\label{fig:scheme}
\end{figure}
Such an approach was used in \cite{PhysRevA.91.053802} to describe SFWM in lossy resonators. We therefore take 
\begin{eqnarray}\label{eq:phantomch}
&\nonumber \mathcal{H}_\text{phantom} = & \int d \omega\, \hbar \omega \bigg[ f^\dagger_{xp}(\omega) f_{xp}(\omega)+\sum_{\nu = i,p,s}f^\dagger_{y\nu}(\omega) f_{y\nu}(\omega)\\
& &+\sum_{\mu = i,s}f^\dagger_{z\mu}(\omega) f_{z\mu}(\omega)\bigg]
\end{eqnarray}
for the external phantom bath and 
\begin{eqnarray}
& \mathcal{H}^\text{phantom}_\text{int} =& \frac{i\hbar}{\sqrt{2\pi}} \int d\omega \big[ \sqrt{\gamma_{xp}}\,x^\dagger_p f_{xp}(\omega)\\
&\nonumber &+\sum_{\nu={i,p,s}} \sqrt{\gamma_{y\nu}}\,y^\dagger_\nu f_{y\nu}(\omega)\\
&\nonumber &+\sum_{\mu=i,s}\sqrt{\gamma_{z\mu}}\,z^\dagger_\mu f_{z\mu}(\omega) + \text{H.c.} \big]
\end{eqnarray}
for the interaction between the microrings and phantom waveguides. Here $\gamma_{qp}$, $\gamma_{qs}$, and $\gamma_{qi}$ ($q=\{x,y,z\}$) are the coupling parameters between the microrings and waveguides that correspond to the linear losses rates in the microring resonators. The nonzero commutation relations read: $[x_p,x^\dag_p]=[y_\nu,y^\dag_\nu]=[z_\mu,z^\dag_\mu]=1$, $[a_\nu(\omega),a^\dag_\nu(\omega')]=\delta(\omega-\omega')$ and $[f_{q\nu}(\omega),f^\dag_{q\nu}(\omega')]=\delta(\omega-\omega')$ ($\nu=\{i,p,s\}$, $\mu=\{i,s\}$).

The Heisenberg--Langevin equations that are derived from the Hamiltonian (\ref{eq:Hamilt}) can be written in the frequency domain as
\begin{equation}\label{Heis_Lang_fourier}
\begin{aligned}
\Big[i\Delta_{p} +\frac{\kappa_p}{2}+\frac{\gamma_{xp}}{2}&\Big] x_p(\omega) +g_p y_p(\omega)\\
&=\sqrt{\kappa_p}\, a_{in;p}(\omega)\\
&\quad+ \sqrt{\gamma_{xp}}\,f_{in;xp}(\omega),\\
\Big[i\Delta_{p} +\frac{\gamma_{yp}}{2}&\Big] y_p(\omega)-g_p x_p(\omega)\\
&=\sqrt{\gamma_{yp}}\,f_{in;yp}(\omega),\\
\Big[i\Delta_{\mu} +\frac{\gamma_{y\mu}}{2}&\Big] y_\mu(\omega)-g_{\mu} z_\mu(\omega)\\
&=\sqrt{\gamma_{y\mu}}\,f_{in;y\mu}(\omega),\\
\Big[i\Delta_{\mu} +\frac{\kappa_{\mu}}{2} +\frac{\gamma_{z\mu}}{2}&\Big] z_\mu(\omega)+g_{\mu} y_\mu(\omega)\\
&=\sqrt{\kappa_{\mu}}\,a_{in;\mu}(\omega)\\
&\quad +\sqrt{\gamma_{z\mu}}\,f_{in;z\mu}(\omega),\\
a_{in;p}(\omega)-a_{out;p}(\omega) &= \sqrt{\kappa_p}\,x_p(\omega),\\
a_{in;\mu}(\omega)-a_{out;\mu}(\omega)&=\sqrt{\kappa_{\mu}}\, z_\mu(\omega),\\
f_{in;xp}(\omega)-f_{out;xp}(\omega)&=\sqrt{\gamma_{xp}}\,x_p(\omega),\\
f_{in;y\nu}(\omega)-f_{out;y\nu}(\omega)&=\sqrt{\gamma_{y\nu}}\,y_\nu(\omega),\\
f_{in;z\mu}(\omega)-f_{out;z\mu}(\omega)&=\sqrt{\gamma_{z\mu}}\,z_\mu(\omega),
\end{aligned}
\end{equation}
where $a_{in;\nu}$ ($a_{out;\nu}$) and $f_{in;q\nu}(\omega)$ ($f_{out;q\nu}(\omega)$) are the annihilation operators for the input (output) fields in the real and phantom waveguides, respectively, $\Delta_{\nu} = \omega_{0\nu}-\omega$, and for all the annihilation operators the Fourier transform is defined as $u(t)=(2\pi)^{-1/2}\int d\omega\, e^{-i\omega t} u(\omega)$. \\

Having derived the basic equations, we are able to find the input-output relations. Let us consider the case when $a_{in;p}\neq 0$, while other input fields are equal to zero, which corresponds to the loading of the pump field. Then from Eqs.~(\ref{Heis_Lang_fourier}) we obtain the input-output relations for the pump field operators
\begin{eqnarray}\label{in-out_pump}
&&y_p(\omega)=\nonumber\\
&&\frac{4g_p\sqrt{\kappa_p}\,a_{in;p}(\omega)}{-4\Delta_{p}^2+4g_p^2+2i\Delta_{p}(\kappa_p+\gamma_{xp}+\gamma_{yp})+(\kappa_p+\gamma_{xp})\gamma_{yp}}\nonumber\\
&&\equiv M_p(\omega) a_{in;p}(\omega).
\end{eqnarray}
Similarly, in the case when only $a_{in;\mu}\neq 0$, which corresponds to the loading of the signal ($\mu=s$) or idler ($\mu=i$) field (and unloading them for the reversed time), we get
\begin{eqnarray}\label{in-out_m}
&&y_\mu(\omega)=\nonumber \\ 
&&\frac{4g_\mu\sqrt{\kappa_\mu}\,a_{in;\mu}(\omega)}{-4\Delta_{\mu}^2+4g_\mu^2+2i\Delta_{\mu}(\kappa_\mu+\gamma_{y\mu}+\gamma_{z\mu})+(\kappa_\mu+\gamma_{z\mu})\gamma_{y\mu}}\nonumber\\
&&\equiv M_\mu(\omega) a_{in;\mu}(\omega).
\end{eqnarray}

\bibliographystyle{apsrev4-1}
\bibliography{bibliography}

\end{document}